\documentclass[10pt]{revtex4}
\usepackage{amssymb}
\usepackage{latexsym}

\begin{document}

\title{Thermodynamic motivations of spherically symmetric static metrics}
\author{H. Moradpour$^1$\footnote{h.moradpour@riaam.ac.ir}, S. Nasirimoghadam$^2$}
\address{$^1$ Research Institute for Astronomy and Astrophysics of Maragha (RIAAM),
P.O. Box 55134-441, Maragha, Iran,\\
$^2$ Department of physics, Sirjan University of Technology, P.O. Box 7618646557,
Sirjan, Iran.}
\begin{abstract}
Bearing the thermodynamic arguments together with the two
definitions of mass in mind, we try to find metrics with spherical
symmetry. We consider the adiabatic condition along with the
Gong-Wang mass, and evaluate the $g_{rr}$ element which points to
a null hypersurface. In addition, we generalize the thermodynamics
laws to this hypersurface to find its temperature and thus the
corresponding surface gravity which enables us to get a relation
for the $g_{tt}$ element. Moreover, we investigate the
mathematical and physical properties of the discovered metric in
the Einstein relativity framework which shows that the primary
mentioned null hypersurface is an event horizon. The obtained
energy-momentum tensor equals the energy-momentum tensor of a
polytropic black hole embedded into an anti-de Sitter background.
We also show that if one considers the Misner-Sharp mass in the
calculations, the Schwarzschild metric will be got. The
relationship between the two mass definitions in each metric is
studied. The results of considering the geometrical surface
gravity are also addressed. Our investigation shows that the
geometrical surface gravity's definition is not always compatible
with the validity of the first law of thermodynamics on the
horizons of spherically symmetric static metrics.
\end{abstract}
\maketitle
\section{Introduction}
Spherically symmetric metrics have vast implications in describing
the world around us. For instance, the FRW metric used to describe
the universe expansion \cite{roos}, moreover, these metrics are
used to get the Tolman-Oppenheimer-Volkoff equations which help us
to model some compact objects such as the Neutron stars and the
white dwarfs \cite{gelen}. Therefore, finding out such solutions
is important from both of the cosmological and gravitational point
of views \cite{solution}. It is also useful to note here that
these metrics may include Black Holes (BHs) which respect certain
rules \cite{pois}. Nowadays, these laws are known as the
thermodynamics laws of BHs \cite{pois,B1,B2,H1,H2}.

Thanks to the Jacobson work \cite{J1}, it seems that the mutual
relation between the BH laws and those of thermodynamics is more
than a simple similarity. Indeed, Jacobson showed that one can
recover the Einstein equations by applying the first law of
thermodynamics on the local Rindler horizon \cite{J1}. Its
generalization to the $f(R)$ gravity which leads to an entropy
production, due to the $R^{n\neq1}$ terms, can also be found in
Ref.~\cite{J11}. In addition, it is shown that one can also
recover the Einstein equations on the event horizon of spherically
symmetric static metrics by applying the first law of
thermodynamics on it \cite{T11}. Generalization to more
gravitational theories are addressed in Ref.~\cite{Pad}. We should
note here that in order to investigate the thermodynamics of
systems, one should define a boundary for the system. Moreover,
since the event horizon is a boundary between the timelike and
spacelike phenomenons and it includes the total mass, it can be
considered as a boundary for the gravitational systems \cite{T11}.
In fact, there are various horizons which can play the role of
boundary in studying the gravitational systems \cite{T11}. The
equivalence between the Friedmann equations and the first law of
thermodynamics is also shown by some authors
\cite{Bak,Cai2,Cai3,CaiKim,W1,W2,Cai33,Shey0,Shey01}.
Additionally, the same deduction is valid in braneworld scenarios
\cite{Shey02,Shey1,C1,Shey2,Shey22,Shey23}. A suitable review on
this topic can be found in Ref.~\cite{Pad0}. Briefly, such
attempts may help us provide a thermodynamic description for
gravity.

As already mentioned, one may reach a thermodynamic origin for
gravity by taking into account the Jacobson works \cite{J1,J11}.
Another approach to get a thermodynamic motivation for the gravity
is proposed by Padmanabhan \cite{Padm}. Recently, Verlinde
proposed a new origin for the gravity leading to emerge the
spacetime \cite{Ver}. In his theory, the tendency of system to
increase its entropy together with the availability of the first
law of thermodynamics on a holographic surface (as a causal
boundary) lead to emerge the spacetime as well as the Einstein
equations and therefore, the first and second laws of
thermodynamics are automatically satisfied in this approach which
attracts more attempts to itself
\cite{Cai4,Cai41,Smolin,Li,Tian,Myung1,Vancea,Modesto,Sheykhi1,BLi,Sheykhi2,Sheykhi21,Sheykhi22,Sheykhi23,Ling,Sheykhi24,Gu,Miao1,other,mann,SMR}.
Loosely speaking, he generalized the thermodynamic laws to the
holographic surfaces to find out a motivation for the spacetime
and gravity as the emerging phenomenons in accordance with
thermodynamic arguments. Finally, it should be reminded that
although all of these approaches try to provide a thermodynamic
interpretation for gravity, their definitions of energy differ
from each other. Moreover, in another attempt, Padmanabhan used
the Komar mass definition, the same as Ref.~\cite{Ver}, as well as
the difference between the surface and bulk degrees of freedom to
get the Friedmann equations which is a thermodynamic motivation
for emerging the FRW metric \cite{padm1,sdh}. Therefore, this
approach may be considered as a thermodynamic motivation for the
FRW metric.

Based on the mentioned attempts it is apparent that a deep mutual
relation between the thermodynamics and gravity is unavoidable
which may lead to a thermodynamic motivation for emerging the
spacetime and thus its metric. Moreover, it is also useful to
investigate the possibility of getting the static spherically
symmetric metrics by considering the thermodynamical arguments. In
order to get a thermodynamical motivation for the Schwarzschild
metric, Zhang et al. have been considered the general form of a
static spherically symmetric metric \cite{zhang1}
\begin{equation}
ds^2=-f(r)dt^2+h(r)dr^2+r^2d\Omega_2^2,
\end{equation}
and used the Misner-Sharp mass definition \cite{ms}
\begin{equation}\label{ms}
M_{ms}=r(1-g^{ab}r_{,a}r_{,b})/2,
\end{equation}
together with the geometrical surface gravity definition
\cite{kodama,hay}. Thereinafter, since an adiabatic system
satisfies the $dM_{ms}=0$ condition which means that the
Misner-Sharp mass is constant, they got
\begin{equation}\label{h1f0}
h(r)=(1-\frac{2M_{ms}}{r})^{-1}.
\end{equation}
Finally, by equating the geometrical surface gravity
\cite{kodama,hay} with ordinary surface gravity \cite{pois}, they
could get the Schwarzschild metric $(f(r)=1-\frac{2M_{ms}}{r}))$.
A nice historical review on the Schwarzschild metric can be found
in ref.~\cite{sch}. Inasmuch as authors have not used the
thermodynamic arguments to get a relation for the surface gravity
in ref.~\cite{zhang1}, their recipe is not a fully thermodynamic
one. But, we should note that their approach is powerful and can
easily be generalized to the Reissner-Nordstrom and Schwarzschild
de-Sitter spacetimes, the Gauss-Bonnet and $F(R)$ theories
\cite{zhang1,zhang2}. Accordingly, we confront two questions.
First, is it possible to find an expression for the surface
gravity by using thermodynamic arguments? Second, what are the
results of considering another definitions of mass?

It has been shown by Gong and coworker that if one takes into
account the Misner-Sharp mass, as a definition for energy in
spherically symmetric spacetimes, then the first law of
thermodynamics is not satisfied on the apparent horizon of FRW
metric in Brans-Dicke, nonlinear and scalar-tensor theories of
gravity \cite{gw}. In addition, they proposed a new definition for
mass in spherically symmetric spacetimes (Gong-Wang) which leads
to the same value as that of Misner-Sharp if one evaluates the
mass confined to the apparent horizon of FRW metric. It is also
shown that, for non-static spherically symmetric metrics
describing an object embedded in the FRW background, the Gong-Wang
definition of mass may lead to the same value as that of the
Misner-Sharp definition for energy \cite{mr}.

Our aim in this paper is finding some spherically symmetric static
metrics by considering thermodynamic arguments. Here, we use the
Gong-Wang as well as the Misner-Sharp definitions of mass and
follow the approach of authors introduced in
Refs.~\cite{zhang1,zhang2} to get the $g_{rr}$ element of metric.
In continue, we try to find an expression for the surface gravity,
which leads to the $g_{tt}$ element, by using the thermodynamics
laws as well as the mutual relation between the surface gravity
and temperature. For the sake of simplicity we take $G=c=\hbar=1$
and the $(-,+,+,+)$ signature throughout this paper.

The paper is organized as follows. In the next section, at first,
by taking into account the spherically symmetric static metrics
along with the Gong-Wang definition of mass, we use the adiabatic
condition to evaluate the $g_{rr}$ element of metric. Moreover, we
try to find the surface gravity and thus the $g_{tt}$ element of
metric by using thermodynamic arguments. Some of the mathematical
properties of the discovered metric are studied. Bearing the
Einstein equations in mind, we also investigate some relativistic
properties of metric. Finally, we repeat our recipe by considering
the Minser-Sharp mass which leads to the Schwarzschild metric.
Moreover, The relation between the Gong-Wang and Misner-Sharp
definitions of mass in the discovered metrics is also addressed.
In addition, the geometrical surface gravity is studied and its
results are compared with other definitions of surface gravity and
thermodynamic considerations. The last section is devoted to a
summary and concluding remarks.

\section{Schwarzschild and Schwarzschild-like solutions}
Consider the
\begin{equation}\label{m}
ds^2=-f(r)dt^2+h(r)dr^2+r^2d\Omega_2^2,
\end{equation}
metric together with the Gong-Wang definition of Mass \cite{gw}
\begin{equation}\label{mf00}
M_{gw}=r\frac{1+g^{ab}r_{,a}r_{,b}}{2}
\end{equation}
leading to
\begin{equation}\label{gw}
M_{gw}=\frac{r}{2}(1+h(r)^{-1}),
\end{equation}
for metric~(\ref{m}). For a spacetime with constant mass
\begin{equation}
dM_{gw}=0,
\end{equation}
which is nothing but the adiabatic condition \cite{zhang1,zhang2},
and therefore
\begin{equation}\label{h1}
h(r)=(\frac{2C}{r}-1)^{-1},
\end{equation}
where $C$ is a constant of integration. Now,
substituting~(\ref{h1}) into~(\ref{mf00}) to get
\begin{equation}\label{mf1}
C=M_{gw},
\end{equation}
which means that, since energy and thus mass are positive, $C$
should meet the $C\geq0$ condition. Finally, we obtain
\begin{equation}\label{h1f}
h(r)=(\frac{2M_{gw}}{r}-1)^{-1}.
\end{equation}
Therefore, we get
\begin{equation}\label{mff}
ds^2=-f(r)dt^2+\frac{1}{\frac{2M_{gw}}{r}-1}dr^2+r^2d\Omega_2^2,
\end{equation}
which is very similar to the Schwarzschild metric. Now, consider
hypersurface $\phi=r-c=0$ with normal vector
$n_{\alpha}=\partial_{\alpha}\phi$, simple calculations yield
\begin{eqnarray}\label{normal}
n_{\alpha}n^{\alpha}=\frac{2M_{gw}}{r}-1,
\end{eqnarray}
proving that $r=2M_{gw}$ is a null hypersurface, and thus it may
be considered as a causal boundary for this spacetime and
therefore, it may also be leaded to an event horizon located at
the $r=2M_{gw}$ radii. In fact, in order to deal with an event
horizon at this radii, metric should change its signature at this
radii in a proper way \cite{pois}, while curvature scalars should
not also be diverged at this radii \cite{pois}.

By using~(\ref{ms}) one gets $M_{ms}(r)=r-M_{gw}$ and thus
$M_{ms}=M_{gw}$ whiles $r=2M_{gw}$, telling us that both of these
mass definitions lead to the same value if we evaluate the mass
confined to the null hypersurface with radii $r=2M_{gw}$. This
mutual consistency between the results of using either the
Gong-Wang or Misner-Sharp masses in order to evaluate the energy
confined to a null hypersurface is in line with the dynamic
situations \cite{gw,mr}. It is useful to note here that the
adiabatic condition together with the Gong-Wang definition of mass
helped us to get an expression for $g_{rr}$ leading to relations
for the Gong-Wang and Misner-Sharp masses. Finally, since the null
hypersurface located at $r=2M_{gw}$ includes the total energy
$E=M_{gw}=M_{ms}$, we consider it as the causal boundary for the
system. Therefore, this hypersurface may carry an entropy in
accordance with Bekenstein's argument \cite{pois,B1}
\begin{eqnarray}\label{ent110}
S_A=\frac{A}{4}=\frac{\int_{r=2M_{gw}} r^2d\Omega}{4}=4\pi
M_{gw}^2.
\end{eqnarray}
Now, using $E=M_{gw}$ together with the Clausius-Clapeyron
relation \cite{calen} to obtain the temperature of the null
hypersurface as
\begin{eqnarray}\label{ent11}
T=\frac{\partial E}{\partial S_A}=\frac{1}{8\pi M_{gw}}.
\end{eqnarray}
Nowadays, $T=\frac{|\kappa|}{2\pi}$ is attributed to various
horizons, as the boundary for system, which are null hypersurfaces
\cite{T11,pois}. Here, since we take into account the null
hypersurface located at $r=2M_{gw}$ as the boundary of system, we
generalize this mutual relation between the surface gravity and
temperature to the mentioned null hypersurface.

Moreover, since Eq.~(\ref{mff}) is a spherically symmetric static
metric, by following the recipe of Refs.~\cite{pois,kodama}, we
get
\begin{eqnarray}\label{s10}
\kappa(r)=\frac{1}{2\sqrt{f(r)h(r)}}|f^{\prime}(r)|,
\end{eqnarray}
where $\kappa$ is the surface gravity. In deriving this equation,
$|f^{\prime}(r)|$ is used instead of $f^{\prime}(r)$ in order to
avoid the negative values for the surface gravity, and thus its
corresponding temperature \cite{pois,T11}. Therefore, we get
\begin{eqnarray}\label{s1}
\kappa=\frac{1}{2\sqrt{(f(r)h(r))_{r=2M_{gw}}}}|f^{\prime}(r)_{r=2M_{gw}}|,
\end{eqnarray}
for the hypersurface located at $r=2M_{gw}$. Finally, by combining
Eq.~(\ref{ent11}) with equation~(\ref{s1}) one gets
\begin{eqnarray}\label{f}
\frac{1}{2\sqrt{(f(r)h(r))_{r=2M_{gw}}}}|f^{\prime}(r)_{r=2M_{gw}}|=\frac{1}{4
M_{gw}}.
\end{eqnarray}
It is easy to check that $f(r)=h(r)^{-1}=\frac{2M_{gw}}{r}-1$ is a
solution for this equation, leading to
\begin{equation}\label{mf}
ds^2=-(\frac{2M_{gw}}{r}-1)dt^2+\frac{1}{\frac{2M_{gw}}{r}-1}dr^2+r^2d\Omega_2^2,
\end{equation}
which is similar to the Schwarzschild metric. Therefore, by using
the thermodynamic arguments we get an expression for the surface
gravity which leads to a relation for the $g_{tt}$ element of
metric. In this metric, radii should meet the $r<2M_{gw}$
condition in order to preserve the $(-,+,+,+)$ signature of
metric. A photon which was emitted at $r_e$ with wavelength
$\lambda_e$ is observed at radii $r_o$ with wave length
$\lambda_o$. For this observed photon the redshift is
\begin{eqnarray}
1+z=\frac{\lambda_o}{\lambda_e}=\sqrt{\frac{\frac{2M_{gw}}{r_o}-1}{\frac{2M_{gw}}{r_e}-1}},
\end{eqnarray}
which signals us that there is no observable photon emitted from
$r_e=2M_{gw}$, because $z\rightarrow \infty$. Moreover, the
redshift of a photon observed at origin ($r_0=0$) diverges,
meaning that one cannot communicate with an observer located at
origin. Therefore, there are two singularities in this metric
located at $r=0$ and $r=2M_{gw}$. For the curvature scalars we get
\begin{eqnarray}
K&=&\frac{16(3m^2+r^2-2rm)}{r^6}, \ \ R= \frac{4}{r^2} \ \
\nonumber\\W&=&\frac{16(3m-r)^2}{3r^6},\ \ R_S=\frac{8}{r^4},
\end{eqnarray}
where $K$ and $R$ are the Kretschmann invariant and the Ricci
scalar, respectively. Moreover, $W$ and $R_S$ also denote the Weyl
and Ricci squares, respectively. It is obvious that none of them
diverges at $r=2M_{gw}$. Finally, based on the above discussions,
we conclude that $r=2M_{gw}$ points to an event horizon which
confirms our primary conjectures about the corresponding
hypersurface including that it may be a causal boundary which may
carry an entropy~(\ref{ent110}) together with a
temperature~(\ref{ent11}) which satisfies Eq.~(\ref{f}). For the
surface area of a sphere with radii $r$ we have
\begin{eqnarray}\label{surfacearea}
A=\int r^2 d\Omega=4\pi r^2,
\end{eqnarray}
leading to $A\rightarrow 0$ whiles $r\rightarrow 0$. It is also
clear that the curvature scalars diverge at origin. Therefore,
based on the above discussions, we think that $r=0$ is a naked
singularity \cite{pois}. It is shown that a naked singularity
leads to time delay as well as gravitational lensing which yields
the magnification of images \cite{komar1,komar2}. In fact, since
the existence of naked singularities is not completely rejected in
physics \cite{roger,roger1,roger2,roger3}, the existence of the
$r=0$ naked singularity in this solution is not bad. Bearing the
Einstein equations in mind, we are going to investigate the
properties of metric~(\ref{mf}) in the Einstein general relativity
frame work. Calculations lead to
\begin{eqnarray}\label{Ein1}
\rho=-G_t^t=\frac{2}{r^2},\ \ P_r=G_r^r=-\frac{2}{r^2},
\end{eqnarray}
where the other elements vanish. $\rho$ and $P_r$ denote the
energy density and redial pressure of a fluid supporting this
spacetime, respectively. If we define the radial state parameter
of supporter fluid as $\omega_r=\frac{P_r}{\rho}$, we get
$\omega_r=-1$ which is similar to the state parameter of
cosmological constant used to explain the current accelerating
phase of the universe \cite{roos}. It is apparent that all of the
energy conditions are marginally satisfied. It is because the
transverse pressures are zero
($G_{\theta}^{\theta}=G_{\phi}^{\phi}=0$) whiles $\rho+P_r=0$ and
$\rho>P_r$. It is useful to note here that the energy-momentum
tensor~(\ref{Ein1}), needed for supporting the
geometry~(\ref{mf}), is equal to the energy-momentum tensor of a
polytropic black hole embedded into the anti-de Sitter background
\cite{SS,SS1}. In fact, since the adiabatic approximation is the
backbone of getting metric~(\ref{mf}), and because an adiabatic
process is indeed a polytropic process \cite{CALEN,cris}, such
resemblance between the obtained energy-momentum
tensor~(\ref{Ein1}) and that of Refs.~\cite{SS,SS1} is reasonable.
The geometrical surface gravity ($\kappa_{g}$) due to an
energy-momentum source with energy $M$ and work density
$w=-\frac{1}{2}(T_0^0+T_1^1)$ is defined as
\cite{zhang1,kodama,hay,zhang2}
\begin{eqnarray}\label{geo}
\kappa_{g}=\frac{M}{r^2}-4\pi r w
\end{eqnarray}
leading to
\begin{eqnarray}\label{geo1}
\kappa_{g}=\frac{M_{gw}}{r^2}-8\frac{\pi}{r},
\end{eqnarray}
for metric~(\ref{mf}). For the event horizon, this relation is
reduced to
\begin{eqnarray}\label{geo2}
\kappa_{g}=\frac{1}{4M_{gw}}-4\frac{\pi}{M_{gw}}.
\end{eqnarray}
It is clear that these results differ from those obtained from
Eqs.~(\ref{s10}),~(\ref{ent11}). Indeed, it is easy to check that
the $\kappa_g=\kappa$ condition does not lead to
$f(r)=\frac{1}{h(r)}$ and thus~(\ref{mf}). It is shown that if one
either uses~(\ref{s10}) or~(\ref{ent11}) as a definition for the
surface gravity or the temperature of horizon, then the first law
of thermodynamics is available on the event horizon of the
spherically symmetric static metrics \cite{J1,T11}. Hence, since
Eq.~(\ref{geo1}) does not yield the same value as those of
Eqs.~(\ref{s10}) and~(\ref{ent11}), the first law of
thermodynamics is not available on the event horizon of
metric~(\ref{mf}) if one uses~(\ref{geo1}) as a suitable
definition for the surface gravity and the corresponding
temperature. The latter means that this definition of surface
gravity is not extendable to every metric such as~(\ref{mf}).
\subsection*{Schwarzschild metric}
By using the Misner-Sharp mass~(\ref{ms}) and following the recipe
used to derive~(\ref{h1f}) and~(\ref{normal}), one gets
\begin{equation}\label{hms}
h(r)=(1-\frac{2M_{ms}}{r})^{-1},
\end{equation}
and
\begin{eqnarray}\label{nor}
n_{\alpha}n^{\alpha}=1-\frac{2M_{ms}}{r},
\end{eqnarray}
respectively. The latter signals that there is a null hypersurface
located at $r=2M_{ms}$. Therefore, metric can be written as
\begin{equation}
ds^2=-f(r)dt^2+\frac{1}{1-\frac{2M_{ms}}{r}}dr^2+r^2d\Omega_2^2.
\end{equation}
Now. let us calculate the Gong-Wang mass for this metric which
yields $M_{gw}(r)=r-M_{ms}$ indicating that both of the
Misner-Sharp and Gong-Wang mass definitions point to the same
value if we evaluate the mass confined to the $r=2M_{ms}$ radii.
Once again, we see that both of these definitions estimate the
same value for the mass confined to the horizon which is in
agreement with attempts in which the dynamic black holes have been
studied \cite{mr,gw}. Therefore, as the previous case, $r=2M_{ms}$
may be considered as a boundary for the system. Finally, in
accordance with~(\ref{nor}), by assuming that the thermodynamics
laws are available on the null hypersurfac located at $r=2M_{ms}$,
and following the recipe used to derive~(\ref{f}), we get
\begin{eqnarray}
\frac{1}{2\sqrt{(f(r)h(r))_{r=2M_{ms}}}}|f^{\prime}(r)_{r=2M_{ms}}|=\frac{1}{4
M_{ms}}.
\end{eqnarray}
It is apparent that $f(r)=h(r)^{-1}$ is a solution for this
equation leading to
\begin{equation}\label{mf3f}
ds^2=-(1-\frac{2M_{ms}}{r})dt^2+\frac{1}{1-\frac{2M_{ms}}{r}}dr^2+r^2d\Omega_2^2,
\end{equation}
which is nothing but the Schwarzschild metric. Previously, Zhang
et al.~\cite{zhang1,zhang2} derived this metric by focusing on the
Misner-Sharp mass~(\ref{ms}) together with the geometrical
definition of surface gravity~(\ref{geo}). Here, bearing the
adiabatic condition in mind, we used the Misner-Sharp mass to get
the $g_{rr}$ element of metric which leads to discover a null
hypersurface. Moreover, the mass analysis guided us to take into
account this null hypersurface as a boundary for the system.
Finally, by establishing the Clausius-Clapeyron relation on the
mentioned null hypersurface and generalizing the mutual relation
between the surface gravity and temperature to the assumed
boundary, we reached the $g_{tt}$ element of metric leading to get
the Schwarzschild metric. We should note here that if one uses the
Misner-Sharp mass then all of the definitions of surface gravity,
introduced in this paper, lead to the same value for the surface
gravity in the Schwarzschild spacetime. It is due to this fact
that the Schwarzschild metric is a vacuum solution which leads to
$w=0$.
\section{Summary and concluding remarks}
Here, we took into account a spherically symmetric static metric
with unknown $g_{tt}$ and $g_{rr}$. Thereinafter, we used the
Gong-Wang definition of mass and applied the adiabatic condition
to the system ($dE=0$) in order to find an expression for $g_{rr}$
which points to a null hypersurface. By this work, we could find a
relation for the Gong-Wang and Misner-Sharp masses. We also saw
that both of these mass definitions lead to the same value for the
mass confined to the discovered null hypersurface located at
$r=2M_{gw}$. In addition, we considered this hypersurface as a
boundary for the system and assigned an entropy and a temperature
to it in accordance with Bekenstein's arguments as well as the
Clausius-Clapeyron relation. In continue, by generalizing the
mutual relation between the temperature of the horizons and their
surface gravity to the boundary of our system we could find an
expression for the corresponding surface gravity. Moreover, we
have estimated the surface gravity of the primary assumed metric
and equated it with the thermodynamic outcomes. The latter leaded
to a relation for the $g_{tt}$ element of metric. Therefore, we
could find a new metric by relying on thermodynamic arguments.
Additionally, by studying the redshift and the mathematical
properties of the metric, we found that the primary null
hypersurface, assumed as the causal boundary, is an event horizon
whiles, there is a naked singularity located at origin ($r=0$).
Bearing the Einstein equations in mind, we investigated the
physics behind this metric including the properties of a fluid
needed for supporting this spacetime and the validity of the
energy conditions. Our study shows that the energy conditions are
marginally satisfied which are due to this fact that $\rho+P_r=0$
whiles, $\rho=\frac{2}{r^2}$ and the other components vanish. As
we have previously mentioned, the obtained energy-momentum tensor
is equal to that of a polytropic black hole embedded into an
anti-de Sitter background. This similarity is due to this fact
that the adiabatic process (the primary assumption used to get the
$g_{rr}$ element of metrics in this paper) is in fact a polytropic
process. Thereinafter, we considered the Misner-Sharp mass, and
followed our hypothetical recipe to get the Schwarzschild metric.
Although some authors gave a thermodynamical motivation for the
Schwarzschild metric in various theories of
gravity~\cite{zhang1,zhang2}, their approach differs from ours.
Here, we focused on the thermodynamic considerations for getting
the surface gravity whiles, they have used the geometrical surface
gravity definition. Moreover, we have also shown that the
geometrical surface gravity is not always compatible with other
definitions of surface gravity and thus the thermodynamic
considerations, such as the validity of the first law of
thermodynamics on the event horizon of spherically symmetric
static metrics. Moreover, our study shows that both of the
Misner-Sharp and Gong-Wang definitions of mass lead to the same
value for the energy confined to the horizon and thus the black
hole mass which is in line with the dynamic situations
\cite{mr,gw}. Finally, we think that our investigation may help us
to provide the thermodynamic motivations for the spherically
symmetric static metrics which leads to clarify the thermodynamic
aspects of spacetime and gravity.
\section*{Acknowledgments}
The work of H. Moradpour has been supported financially by
Research Institute for Astronomy \& Astrophysics of Maragha
(RIAAM) under research project No. $1/4165-4$

\end{document}